\documentclass[a4paper,11pt]{article}
\usepackage{pos}

\title{Modified Hadronic Interactions and the future of UHECR observations}
\author*[a]{Jan Ebr}
\author[a]{Jiří Blažek}
\author[a]{Jakub Vícha}
\author[b]{Tanguy Pierog}
\author[a]{Eva Santos}
\author[a]{Petr Trávníček}
\author[a]{Nikolas Denner}
\author{Ralf Ulrich}

\affiliation[a]{FZU - Institute of Physics of the Czech Academy of Sciences,\\
 Na Slovance 2, Prague, Czech Republic}

\affiliation[b]{Karlsruhe Institute of Technology\\
76131 Karlsruhe, Germany}

\emailAdd{ebr@fzu.cz}

\abstract{Data from multiple experiments suggest that the current interaction models used in Monte Carlo simulations do not correctly reproduce the hadronic interactions in air showers produced by ultra-high-energy cosmic rays (UHECR). We have created a large library of UHECR simulations where the interactions at the highest energies are slightly modified in various ways – but always within the constraints of the accelerator data, without any abrupt changes with energy and without assuming any specific mechanism or dramatically new physics at the ultra-high energies. Recent results of the Pierre Auger Observatory indicate a need for a change in the prediction of the models for both the muon content at ground and the depth of the maximum of longitudinal development of the shower. In our parameter space, we find combinations of modifications that are in agreement with this analysis, however a consistent description of UHECR showers remains elusive. Our library however provides a realistic representation of the freedom in the modeling of the hadronic interactions and offers an opportunity to quantify uncertainties of various predictions. This can be particularly valuable for the design of future observatories where hadronic models are often used as input for the prediction of the performance. We demonstrate this powerful capability on several selected examples.}

\FullConference{7th International Symposium on Ultra High Energy Cosmic Rays (UHECR2024)\\
 17-21 November 2024\\
Malargüe, Mendoza, Argentina\\}

\begin{document}
\maketitle

\section{Modified hadronic interactions: summary}

In several previous works \cite{icrc2021,uhecr2022,icrc2023,rumunsko}, we have described in detail our systematic exploration of the parameter space of ad-hoc modifications of basic parameters UHECR hadronic interactions and of the impact of such modifications of air-shower observables. In our macroscopic approach, we follow the ideas outlined in \cite{ulrich} and modify elasticity ($\eta$), multiplicity ($N$) and cross-section ($\sigma$) of hadronic interactions in an energy-dependent manner: the modifications take effect at a given energy threshold and their magnitude increases logarithmically with energy so that it reaches a set value $f_{19}$ at $10^{19}$ eV. The thresholds and the values for $f_{19}$ are chosen so that the predictions at energies accessible to accelerators stay within the uncertainties of the current measurements. Unlike in \cite{ulrich}, we take advantage of a fully 3-dimensional simulation in CORSIKA \cite{corsika}, which allows us to make direct comparison with the data from UHECR experiments, in particular the Pierre Auger Observatory \cite{auger}. Overall we consider 75 combinations of modifications for two primaries (proton and iron) at a single primary energy $10^{18.7}$ eV and five discrete values of zenith angle between 0 and 60 degrees; with 1000 showers per a combination of input parameters, the library has 750 thousand showers.

The latest analysis of the Pierre Auger Observatory \cite{kuba} indicates that the none of current hadronic interaction models describe the observed UHECR air showers correctly and that changes in the predictions for both the number of muons at ground (at 1000 meters from the shower core) and the depth of the shower maximum ($X_\mathrm{max}$) are needed. Confronting these results with our simulation library, we find that the hadronic model Sibyll 2.3d \cite{sibyll}, which we use as a baseline, can indeed be modified to describe this particular Auger data well within our set parameter space, but only with changes to all three parameters -- a decrease in $\sigma$ and increases in both $N$ and $\eta$. However such changes, in particular the increase in elasticity, are difficult to reconcile with other Auger measurements. The measurement of proton-air cross-section \cite{augercs} using the slope of the exponential tail of the $X_\mathrm{max}$ distribution is effectively a constraint in the $\sigma$-$\eta$ space such that if  $\sigma$ is decreased, then $\eta$ should be also decreased to maintain the observed value of this slope (which is in a good agreement with unmodified Sibyll 2.3d). Furthermore, increasing $\eta$ leads to a marked increase in $X_\mathrm{max}$ fluctuations for proton primaries, which may be at odds with Auger measurements \cite{xmax} unless the primary composition is consistently heavy (detailed analysis of the significance of this tension would however require performing modified simulations with more primary energies and particle types). 

\begin{figure}
\centering
\includegraphics[width=\textwidth]{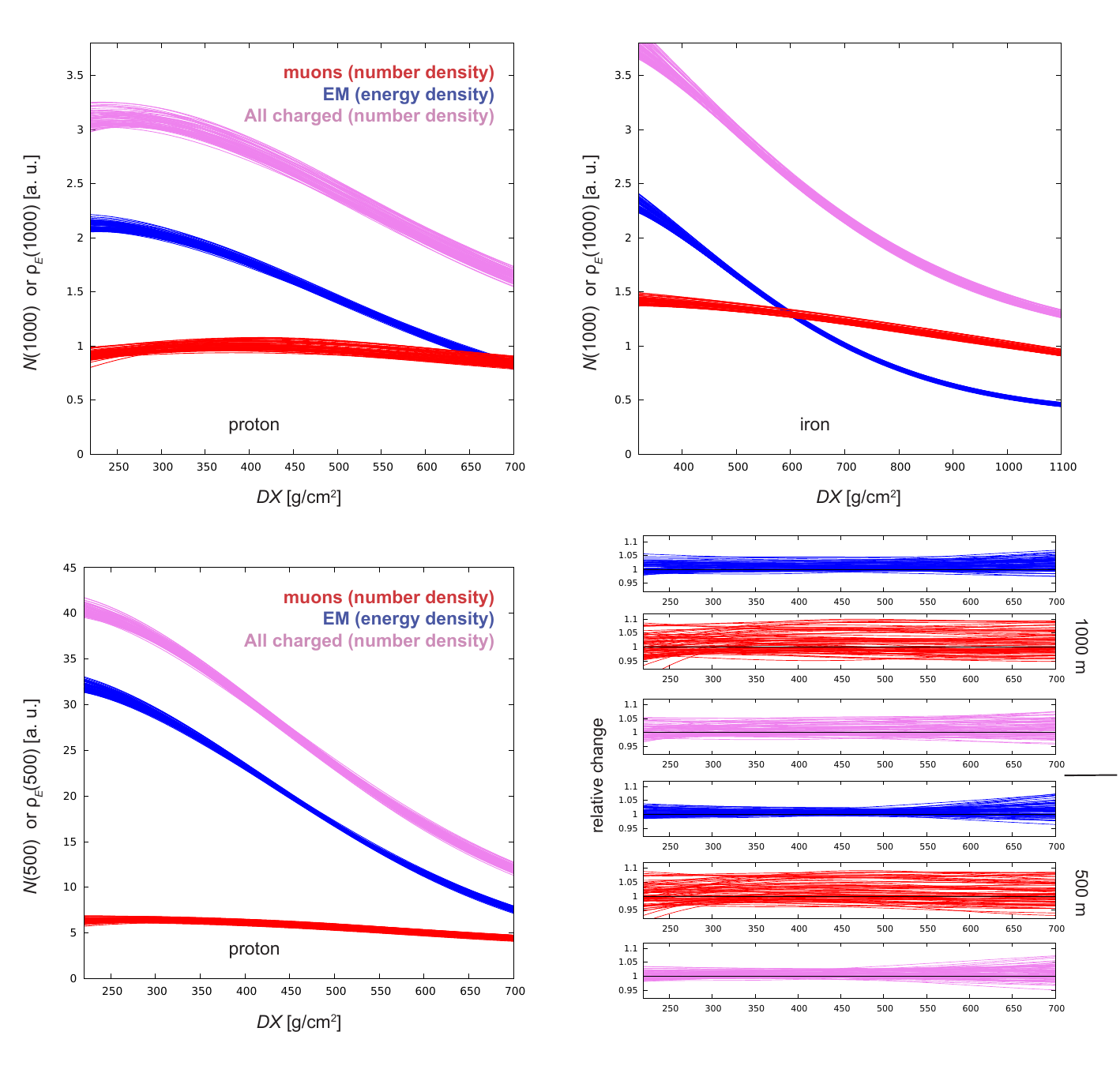}

\caption{D$X$ profiles for the muon number density, electromagnetic energy density and charged-particle number density for proton primaries at 1000 meters from the shower axis (top left), iron at 1000 meters (top right) and proton at 500 meters (bottom left). For each quantity, 75 mean profiles are plotted, each corresponding to a different combination of modifications. Bottom right: relative changes in the profiles with respect to unmodified simulations, for primary  protons.}
\label{fig:unc}
\end{figure}

\section{Uncertainty in the modeling of hadronic interactions}

The differences between the predictions of different hadronic interaction models, currently usually Sibyll 2.3d, EPOS-LHC \cite{eposlhc} and QGSJET-II-04 \cite{qgsii}, are often used as an estimate of the uncertainty in various prediction due to the unknown extrapolation of the accelerator data into the UHECR energy range. However already the aforementioned Auger analysis shows that to describe the air showers properly, all three models may need to be modified in the same direction and thus the "truth" does not necessarily have to be bracketed by the current models. It can thus be beneficial to consider a whole class of modified simulations when estimating the modeling uncertainty. 

As an illustration of this concept, let us consider that for a general UHECR observatory, the detectors on the ground will typically  fall into one of three categories - the measured signal will be proportional to either the number density of muons, the number density of any charged particles or the energy density of EM particles, all at some fixed distance from the shower core, the choice of which is determined by the geometry of the detector. The predictions for these quantities vary significantly between different modifications, but a large part of these variations is simply due to the changes in $X_\mathrm{max}$  which, for our set of parameters, range between $-45$  and $+55$ g/cm$^2$. When the predicted ground signal is expressed as a function of $\mathrm{D}X=X-X_\mathrm{max}$, we find that the universality of shower profiles \cite{universality} is well preserved. 

The shower-to-shower fluctuations are generally larger than the differences between individual modifications. Thus, for each combination of modifications and primary particle, we fit any chosen quantity at ground as a function of D$X$ (in the form of a Gaisser-Hillas profile with an additional constant term) to all showers simulated at the higher three of the five zenith angles (38, 49 and 60 degrees), where the maximum is always sufficiently above ground and thus precise determination of $X_\mathrm{max}$ is possible. For observables at 1000 meters, this allows for stable fits for for D$X>200$ g/cm$^2$ (Fig.~\ref{fig:unc}). We find that for primary protons at 500 meters from the shower axis, the EM energy density and the charged-particle number density at a given D$X$ are both conserved within a few percent, while the number of muons can be increased by up to 10 \% by some modifications (interestingly, it is never decreased to such an extent). At 1000 meters, the variations in the number of muons are roughly the same, but the variations in the other two quantities increase, more so for the charged-particle number density. This example shows the importance of the 3-D approach in quantifying the model uncertainties.

\section{Proton–iron separation}

The Sibyll interaction model implements nucleus-air interactions as a pure superposition of nucleon-air interactions and we thus implement the modifications for nuclear projectiles simply by modifying the underlying nucleon-air interactions. To do so consistently, we must consider each of these interactions at 1/A of the total primary energy -- this however decreases the modification factor applied by bringing the primary energy closer to the respective thresholds. Thus, the effects of modifications for iron primaries are always smaller than for proton primaries. Because the number of muons is higher in iron showers, this means that when the number of muons is increased, the difference between iron and proton showers decreases (and vice versa). 

\begin{figure}
\centering
\includegraphics[width=\textwidth]{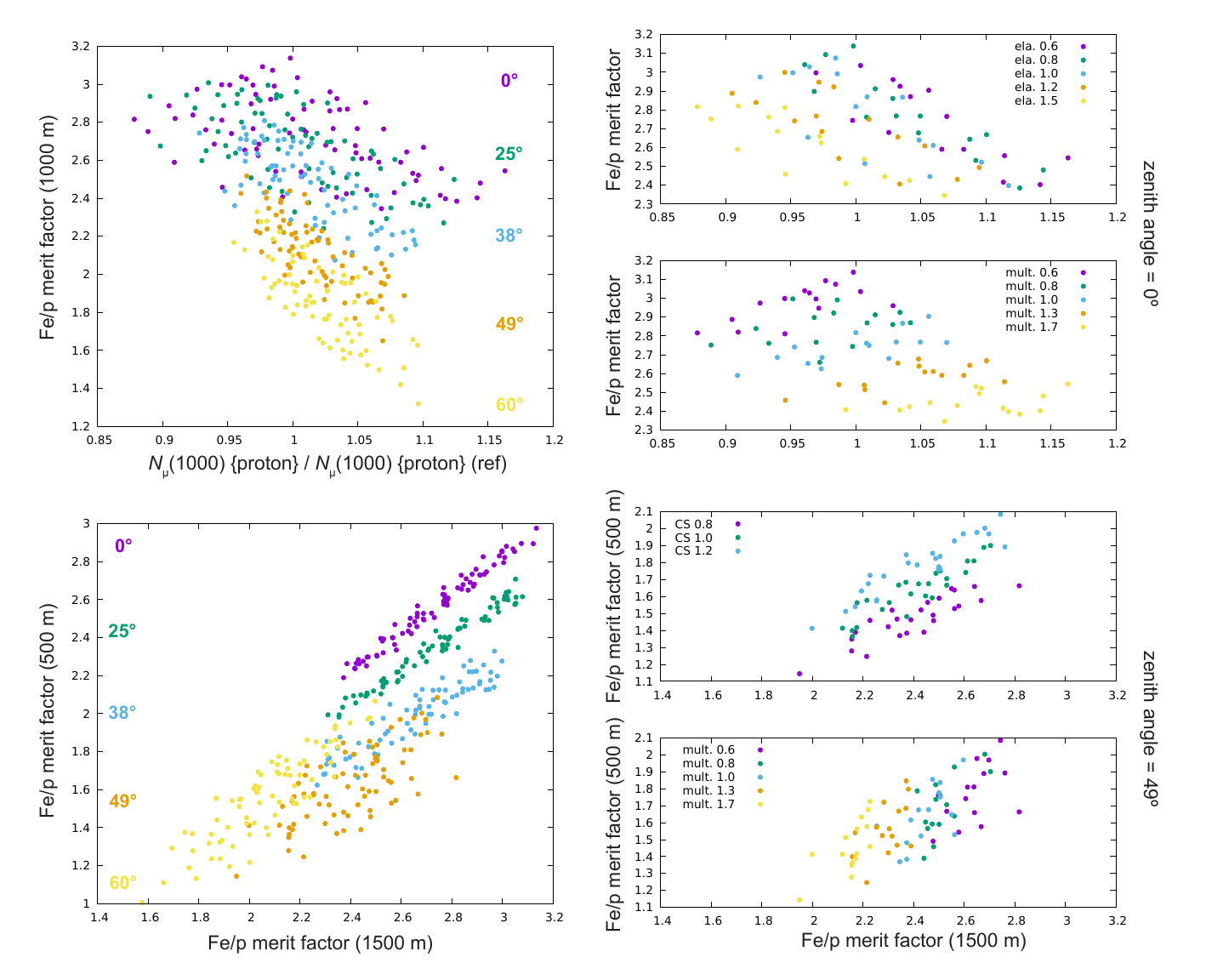}
\caption{Top left: Correlation between the merit factor for iron-proton separation with muons at 1000 meters from the shower axis and the change in the number of muons for proton primaries at different zenith angles (color coded). Top right: points for vertical showers colored by changes in $\eta$ and $N$. Bottom left: Correlation between merit factors for iron-proton separations with muons at 500 and 1500 meters from the shower axis at different zenith angles (color coded). Bottom right: points for showers at 49 degrees colored by changes in $\sigma$  and $N$.}
\label{fig:mumer}
\end{figure}

To quantify the ability of an observatory to separate between proton and iron using an general observable $S$, it is practical to use the merit factor $MF$, defined as

\begin{equation}
    MF=\frac{\langle S_p\rangle-\langle S_{Fe}\rangle}{\sqrt{\sigma_{S_p}^2+\sigma_{S_{Fe}}^2}}.
\end{equation} 

While the ratio between the number of muons for iron and proton at 1000 m from the shower axis is very well correlated with the increase in the number of muons for proton, the merit factor of the same variable shows such strong correlation only for large zenith angles. Looking more into the 3-D features of the problem, we find that the iron/proton merit factors for muons at 500 m and 1500 m are tightly correlated for small zenith angles, but less so for large zenith angles. At 49 degrees, we can see that the changes in the 500-m merit factor are dominated by modifications of $\sigma$, while the 1500-m depends mostly on $N$ (Fig.~\ref{fig:mumer}). Observations like this may have important consequences for the choice of optimal observatory design, in particular as the ranges of allowed parameters for different modifications will be increasingly constrained in the future.

Finally in this topic, it is interesting to consider that what most observatories measure is not purely "the number of muons", but "the number of muons observed at a primary energy established by other means". The simplest way to take this into account is to consider ratios between observables of different types of detectors instead of just the number of muons. At 1000 meters, the three possible ratios between the three "types of detectors" previously discussed (muon number, charged number and EM density) produce iron/proton merit factors that are strongly correlated, yet with non-zero intercepts for low zenith angles; at 500 meters, the picture is much more complex.

\begin{figure}
\centering
\includegraphics[width=0.99\textwidth]{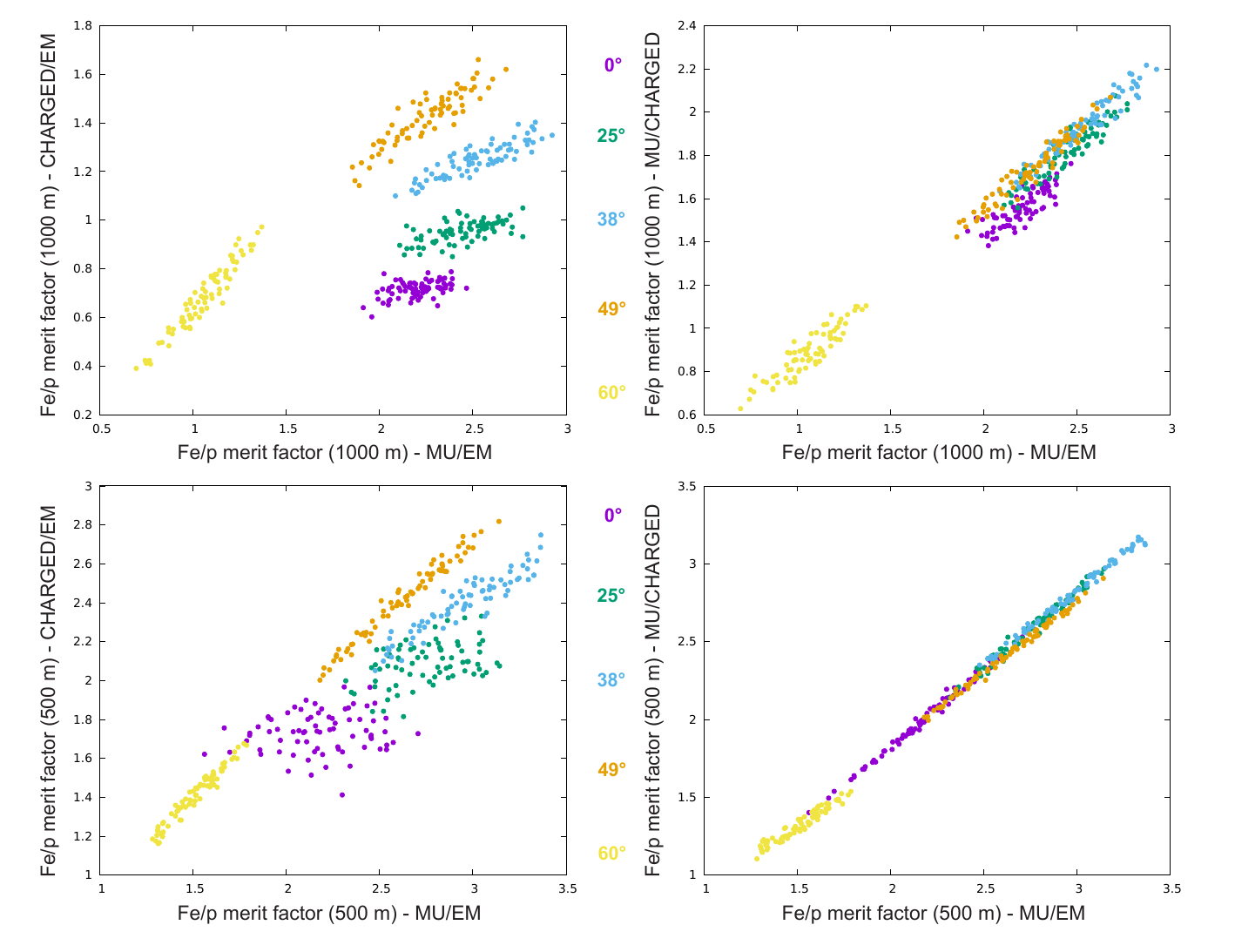}
\caption{Left column: comparison between merit factors for iron-proton separation using the ratio of number density of muons to energy density of electromagnetic particles and using the ratio of number density of all charged particles to energy density of electromagnetic particles. Right column: comparison between merit factors for iron-proton separation using the ratio of number density of muons to energy density of electromagnetic particles and using the ratio of number density of muons to that  of all charged particles. Top row at 1000 meters from the shower axis, bottom row at 500 meters. Color coded for zenith angle. }
\label{fig:difobs}
\end{figure}

It is important to note that all the observed correlations are a direct consequence not only of the choice to implement modified nucleus-air interactions as a superposition of modified nucleon-air interactions, but also of the gradual rise of the modifications from their respective thresholds. While both of these choices are in many aspects natural, we certainly do not claim that other possibilities should not be considered.

\section{Conclusions}

Changing cross-section, elasticity and multiplicity within reasonable limits can have major impact on
air-shower properties, and this impact can be quite different for quantities depending on 3D geometry as opposed to 1D sums. The changes of hadronic interactions indicated by the Pierre Auger Observatory are just reachable, but only with a combination of modifications and already in a tension with other measurements.  Even if some modifications are not realistic, we can learn interesting insights from the extensive shower library produced. We find that the effects of 3D modifications are highly dependent on the distance to shower axis, that the number of muons is more affected than the EM energy density and that proton/iron separation power can vary significantly and with a complex dependence on the type of detector and its geometry (but note the implicit assumption on the implementation of modifications for nuclear primaries). The gamut of 75 slightly different realizations of hadronic interactions offers an interesting alternative to just using the three major models for the estimation of any systematic uncertainties.

\acknowledgments

This work was supported by the following sources: Czech Academy of Sciences: LQ100102401, Czech Science Foundation: 21-02226M, Regional funds of EU/MEYS: OPJAK FORTE CZ.02. 01.01/00/22\_008/0004632.

\bibliographystyle{JHEP}
\bibliography{bibi}

\end{document}